\documentclass[10pt, conference, compsocconf]{IEEEtran}
\usepackage{times}

\newif\ifdraft\drafttrue
\newif\ifnotes\notesfalse

\usepackage[stretch=10,shrink=10]{microtype}
\usepackage{mathptmx}
\usepackage{ptmgreek}
\usepackage{framed}
\usepackage{graphicx}
\usepackage{booktabs}
\usepackage{multirow}
\usepackage{ifthen}
\usepackage{fancyhdr}
\usepackage{layout}
\usepackage[end]{algpseudocode}
\usepackage[usenames,dvipsnames,svgnames,table]{xcolor}

\makeatletter
\newif\if@restonecol
\makeatother

\usepackage{hyperref}
\usepackage{algorithm}
\usepackage{url}
\usepackage{mdwlist}
\usepackage[textfont={bf}, labelfont=bf]{caption}
\usepackage[margin=-25pt,format=hang]{subfig}
\usepackage{epsfig}
\usepackage{amssymb}
\usepackage{amsmath}
\usepackage{amsfonts}
\usepackage{xspace}

\newcommand{\eat}[1]{}

\ifdraft\else\fi

\pdfpageattr {/Group << /S /Transparency /I true /CS /DeviceRGB>>}

\ifnotes
\newenvironment{notes}[1][]{\begin{shaded}}{\end{shaded}}
\else
\fi

\def\urltilda{\lower .7ex\hbox{\~{}}\kern .04em}

\ifnotes
\newcommand{\debug}[1]{{\bf [#1]}}
\else
\newcommand{\debug}[1]{}
\fi

\usepackage{tikz}
\usetikzlibrary{arrows, positioning, shadows, plotmarks}
\pgfdeclarelayer{background}
\pgfdeclarelayer{foreground}
\pgfsetlayers{background,main,foreground}
 
\makeatletter
  \let\@copyrightspace\relax
  \makeatother

\newcommand{\algoname}{CloudPowerCap\xspace}

\begin{document}

\title{CloudPowerCap: Integrating Power Budget and Resource Management
across a Virtualized Server Cluster}
\date{}

\author{Yong Fu\IEEEauthorrefmark{1}, Anne
Holler\IEEEauthorrefmark{2}, Chenyang Lu\IEEEauthorrefmark{1} \\
{\small \IEEEauthorrefmark{1} Computer Science and Engieering Department, Washington University in
St. Louis} \\ 
{\small \IEEEauthorrefmark{2} Distributed
Resource Management Team, VMware}  }

%
%

\maketitle


\IEEEaftertitletext{\vspace{-3\baselineskip}}
\begin{abstract}
In many datacenters, server racks are highly underutilized.  Rack
slots are left empty to keep the sum of the server nameplate maximum
power below the power provisioned to the rack.  And the servers that are
placed in the rack cannot make full use of available rack power.  The
root cause of this rack underutilization is that the server nameplate
power is often much higher than can be reached in practice.  To
address rack underutilization, server vendors are shipping
support for per-host power caps, which provide a server-enforced limit
on the amount of power that the server can draw.  Using this feature,
datacenter operators can set power caps on the hosts in the rack to
ensure that the sum of those caps does not exceed the rack's
provisioned power.  While this approach improves rack utilization, it
burdens the operator with managing the rack power budget across the
hosts and does not lend itself to flexible allocation of power to
handle workload usage spikes or to respond to changes in the amount of
powered-on server capacity in the rack.  In this paper we present
CloudPowerCap, a practical and scalable solution for power budget
management in a virtualized environment.  CloudPowerCap manages the
power budget for a cluster of virtualized servers, dynamically
adjusting the per-host power caps for hosts in the cluster.  We show
how CloudPowerCap can provide better use of power than
per-host static settings, while respecting virtual machine resource
entitlements and constraints.


\end{abstract}

\begin{IEEEkeywords}
power cap; resource management; virtualization; cloud computing
\end{IEEEkeywords}

\section{Introduction}
\label{sec:intro}

In many datacenters, server racks are as much as 40 percent
underutilized \cite{gartner-rack-low-util}.  Rack slots are
intentionally left empty to keep the sum of the servers' nameplate
power below the power provisioned to the rack.  And the servers that
are placed in the rack cannot make full use of the rack's provisioned
power.  The root cause of this rack underutilization is that a
server's peak power consumption is in practice often significantly
lower than its nameplate power \cite{Fan:2007:PPW}. This server rack
underutilization can incur substantial costs.  In hosting facilities
charging a fixed price per rack, which includes a power charge that
assumes the rack's provisioned power is fully consumed, paying a 40
percent overhead for rack underutilization is nontrivial.  And in a
private datacenter, the amortized capital costs for the infrastructure
to deliver both the racks' provisioned power and the cooling capacity
to handle the racks' fully populated state comprises 18 percent of a
datacenter's total monthly costs \cite{fullburden}.  If that
infrastructure is 40 percent underutilized, then 7 percent of the data
center's monthly costs are wasted for this reason.

Due to the significant cost of rack underutilization, major server
vendors are now shipping support for per-host power caps, which
provide a hardware or firmware-enforced limit on the amount of power
that the server can draw~\cite{HPDPCHowItWorks,DELLESM,IBMAEM}.  These
caps work by changing processor power states~\cite{acpi} or by using
processor clock throttling, which is effective since the processor is
the largest consumer of power in a server and its activity is highly
correlated with the server's dynamic power
consumption~\cite{Fan:2007:PPW, HPDPCHowItWorks}.  Using per-host
power caps, data center operators can set the caps on the servers in
the rack to ensure that the sum of those caps does not exceed the
rack's provisioned power.  While this approach improves rack
utilization, it burdens the operator with manually managing the rack
power budget allocated to each host in a rack.  In addition, it does
not lend itself to flexible allocation of power to handle workload
spikes or to respond to the addition or removal of a rack's powered-on
server capacity.

Many datacenters use their racked servers to run virtual machines
(VMs).  Several research projects have investigated power cap
management for virtualized infrastructure~\cite{Raghavendra2008,
  Nathuji2009, Nathuji2009a, Lim2011, Wang2011, Davis2012}.  While
this prior work has considered some aspects of VM Quality-of-Service
(QoS) in allocating the power budget, it has not explored operating in
a coordinated fashion with a comprehensive resource management system
for virtualized infrastructure.  Sophisticated cloud resource
management systems such as VMware Distributed Resource Scheduler (DRS)
support admission-controlled resource reservations, resource
entitlements based fair-share scheduling, load-balancing to maintain
resource headroom for demand bursts, and respect for
constraints to handle user's business
rules~\cite{Gulati2011}.  The operation of virtualized infrastructure
resource management can be compromised if power cap budget management
is not tightly coordinated with it.
\begin{itemize} \itemsep1pt \parskip0pt \parsep0pt \listparindent0pt
\item Host power cap changes may cause the violation of
VMs' resource reservations, impacting end-users' Service-Level
Agreements (SLAs).
\item Host power cap changes may interfere with
the delivery of VMs' resource entitlements, impacting resource
fairness among VMs.
\item Host power cap changes may lead to
imbalanced resource headroom across hosts, impacting peak performance
and robustness in accommodating VM demand bursts.
\item Power cap settings may limit the ability of the infrastructure to
respect constraints, impacting infrastructure usability.
\item For resource  management systems supporting
power proportionality via powering hosts off and on along
with changing the level of VM consolidation, host power cap settings may
cause the power budget to be inefficiently allocated to hosts,
impacting the amount of powered-on computing capacity available for a
given power budget.
\end{itemize}

This paper presents CloudPowerCap, an autonomic computing approach to power 
budget management in a virtualized environment. CloudPowerCap manages the power 
budget for a cluster of virtualized servers, dynamically adjusting the per-host power 
caps for servers in the cluster. It allocates the power budget in close coordination with a 
cloud resource management system, operating in a manner consistent with the system’s 
resource management constraints and goals. To facilitate interoperability between power 
cap and resource management, CloudPowerCap maps a server’s power cap to its CPU 
capacity and coordinate with the resource management system through well defined 
interfaces and protocols. The integration of power cap and resource management 
results in the following novel capabilities in cloud management.
\begin{itemize}
\item \textbf{Constraint satisfaction via power cap reallocation:} Dynamic power cap reallocation 
enhances the system’s capability to satisfy VM constraints, including resource 
reservations and business rules.
\item \textbf{Power-cap-based entitlement balancing:} Power cap redistribution provides an 
efficient mechanism to achieve entitlement balancing among servers to provide 
fairness in terms of robustness to accommodate demand fluctuation. Power-cap-
based entitlement balancing can reduce or eliminate the need for moving VMs for 
load balancing, reducing the associated VM migration overhead.
\item \textbf{Power cap redistribution for power management:} CloudPowerCap can redistribute 
power caps among servers to handle server power-off/on state changes caused by 
dynamic power management. Power cap redistribution reallocates the power budget 
freed up by powered-off hosts, while reclaiming budget to power-on those hosts 
when needed.
\end{itemize}
We have implemented and integrated CloudPowerCap with VMware Distributed
Resource Scheduler (DRS). Evaluation based on an industrial cloud simulator
demonstrated the efficacy of integrated power budget and resource management in
virtualized servers clusters.

\section{Motivation}
\label{sec:motive}

In this section, we motivate the problem \algoname is intended to
solve.  We first describe the power model mapping a host's power cap to
its CPU capacity, which enables CloudPowerCap to integrate power cap
management with resource management in a coordinated fashion. We next
discuss some trade-offs in managing a rack power budget.  After a
brief introduction of the resource management model, we then provide
several examples of the value of combining dynamic rack power budget
management with a cloud resource management system. 

\subsection{\algoname Power Model}
\label{sec:power_model}

The power model adopted by \algoname maps the power cap of the host to
the CPU capacity of the host, which is in turn managed by a resource
management system directly.  A host's power consumption $P_{consumed}$
is commonly estimated by its CPU utilization $U$ and the idle
$P_{idle}$ and peak $P_{peak}$ power consumption of the host via a
linear function, which is validated by real-world workloads in
previous measurements and analysis~\cite{intelpower, Fan:2007:PPW},
\setlength{\abovedisplayskip}{0pt}
\setlength{\belowdisplayskip}{0pt}
\begin{equation}\label{eq:linpower}
  P_{consumed} = P_{idle} + (P_{peak} - P_{idle}) U.
\end{equation}
The power $P_{idle}$ represents the power consumption of the host when
the CPU is idle.   $P_{idle}$ intentionally
includes the power consumption of the non-CPU components, such as
spinning disk, since in enterprise datacenter shared storage is
usually employed and their power draw does not vary significantly with
utilization.  The power $P_{peak}$ represents the power consumption of
the host when the CPU is $100\%$ utilized at its maximum CPU capacity
$C_{peak}$, with the CPU utilization $U$ expressed as a fraction of
the maximum capacity.

We note that the power estimate $P_{consumed}$ is an upper-bound if a
host power management technology such as dynamic voltage and frequency
scaling (DVFS) is used.  DVFS can deliver a percentage of maximum CPU
capacity at a lower power consumption, e.g., DVFS could deliver
the equivalent of 50 percent utilization of a 2 GHz maximum capacity
processor at lower power consumption by running the processor at 1 GHz
with 100 percent utilization.  Computing $P_{consumed}$ as an upper
bound is desirable for the resource management use case, to ensure
sufficient power budget for worst case.

For a host power cap $P_{cap}$ set below $P_{peak}$,
Equation~\eqref{eq:linpower} can be used to solve for the lower-bound of the CPU
capacity $C_{capped}$ reachable at that power cap, i.e., the host's
effective CPU capacity limit which we refer to as its \emph{power-capped capacity}.
In this case, we rewrite Equation~\eqref{eq:linpower} as:
\begin{equation}
  P_{cap} = P_{idle} + (P_{peak} - P_{idle}) (C_{capped} / C_{peak}).
\end{equation}
and then solve for $C_{capped}$ as:
\begin{equation}\label{eq:power_model}
 C_{capped} = C_{peak} (P_{cap} - P_{idle}) / (P_{peak} - P_{idle}).
\end{equation}

\subsection{Managing a Rack Power Budget}
\label{sec:rackpower}

To illustrate some trade-offs in managing a rack power budget, we
consider the case of a rack with a budget of 8 KWatt, to be populated
by a set of servers.  Each server has 34.8 GHz CPU capacity comprising
12 CPUs, each running at 2.9 GHz, along with the other parameters
shown in Table~\ref{tab:server}. 
\begin{table}[htbp]
  \centering
  {\small 
  \begin{tabular}{@{}ccccc@{}}
    \toprule
    \textbf {CPU}  & \textbf {Memory}  & \textbf {Nameplate} & \textbf {Peak} & \textbf{ Idle} \\
    \midrule
    34.8 GHz & 96 GB  & 400 W & 320 W & 160 W \\
    \bottomrule
  \end{tabular}
}
  \caption{The configuration of the server in the rack.}
  \label{tab:server}
\end{table}

Given the power model presented in the previous section and
the servers in Table~\ref{tab:server},
the rack's 8 KWatt power budget can accommodate various deployments
including those shown in Table~\ref{table:serverPower}.  Based on the 400 Watts
nameplate power, only 20 servers can be placed in the rack.  Instead
setting each server's power cap to its peak attainable power draw of 320 Watts
allows 25 percent more servers to be placed in the rack.  This choice
maximizes the amount of CPU capacity available for the rack power
budget, since it best amortizes the overhead of the servers'
powered-on idle power consumption.  However, if memory may sometimes
become the more constrained resource, the memory made available by
placing additional servers in the rack may be critical.  Setting each
server's power cap to (say) 250 Watts allows 32 hosts to be placed in the
rack, significantly increasing the memory available for the given
power budget.  Note that the additional hosts may also be desirable
in use cases in which a constraint on the number of powered-on VMs per
host has been set to limit the workload impact of a single host failure.

By dynamically managing the host power cap values,
\algoname allows the kinds of trade-offs between CPU and memory capacity illustrated
in Table~\ref{table:serverPower} to be made
at runtime according to the VMs' needs.

\begin{table}[htbp]
\centering
{\small
\begin{tabular}{@{}p{0.35in}p{0.35in}ccccc@{}}
\toprule
\multirow{2}{0.75in}{\textbf{Power Cap(W)}}     & \multirow{2}{*}{\textbf{Count}}  &
\multicolumn{2}{c}{\textbf{CPU}}&\phantom{a} & \multicolumn{2}{c}{\textbf{Memory}}
\\
\cmidrule{3-4}  \cmidrule{6-7}
     &   & \textbf{Capa(GHz)} &\textbf{Ratio} & & \textbf{Size(GB)}   & \textbf{Ratio}  \\
\midrule
400 & 20 & 696 & 1.00 & &1920 & 1.00 \\
320 & 25 & 870 & 1.25 & &2400 & 1.25 \\
285 & 28 & 761 & 1.09 & &2688 & 1.40 \\
250 & 32 & 626 & 0.90 & &3072 & 1.60 \\ 
\bottomrule
\end{tabular}
}
\caption{Server deployments in a rack with 8 KWatt power budget with
  different power caps}
\label{table:serverPower}
\end{table}

\subsection{Resource Management Model}
\label{sec:resmgmt}

The comprehensive resource management system with which \algoname
is designed to interoperate computes each VM's entitled resources
and handles the ongoing location of VMs on hosts so that
the VMs' entitlements can be delivered while respecting constraints,
providing resource headroom for demand bursts, and optionally reducing
power consumption.

\algoname interoperates with support for the following kinds of resource controls,
used to express allocation in terms of \emph{guaranteed service-rate} and/or \emph{relative
importance} (assuming a mapping between service level and delivered resources).
\begin{itemize} \itemsep1pt \parskip0pt \parsep0pt \listparindent0pt
\item \textbf{Reservation:} A reservation specifies the minimum amount of CPU or memory
  resources guaranteed to a VM, even if the cluster is over-committed.  This control is
  expressed in absolute units (e.g., MHz or MB).
\item \textbf{Limit:} A limit specifies the upper bound of CPU or memory resources allocated
  to a VM, even if the cluster is under-committed.  This control is also expressed in absolute units.
\item \textbf{Shares:} Shares specify \emph{relative importance} and
  represent weights of resource allocation used if there is resource contention.
\end{itemize}

Each VM's CPU and memory resource entitlement is computed according to its configuration
and resource control settings, along with an estimate of its CPU and memory
resource \emph{demand}, a metric expressed in absolute units that estimates the amount
of CPU and memory the VM would use to satisfy its workload if there were no contention.
To clarify the CPU \emph{entitlement} model, a VM's entitlement indicates the amount
of CPU capacity the VM deserves to be given by the hypervisor over time in a shared
environment (assuming homogeneous hosts in the cluster).
To illustrate, if a server has an CPU capacity of $4$ GHz,
it can (for example) accomodate 2 VMs, each with an entitlement of 2 GHz.

\algoname interoperates with the following kinds of operations to manage the ongoing location of VMs.
\begin{itemize} \itemsep1pt \parskip0pt \parsep0pt \listparindent0pt
\item \textbf{VM Placement:} VM placement involves initial placement
    of VMs for power-on and relocation of VMs for constraint
    correction to respect user-defined business rules. During initial
    placement, hypervisor hosts are selected to accomodate powering-on VMs.
    User-defined business rules restrict VMs' locations on physical hosts.
\item \textbf{Entitlement Balancing:} Entitlement balancing responds to entitlement imbalance
    by migrating VMs between hosts to avoid potential
    bottlenecks and ensure fairness on performance.
\item \textbf{Distributed Power Management:} To optionally reduce power consumption, the VMs
    distributed across hosts may be consolidated on a subset of the hosts, with the
    vacated hosts powered-off.  Powered-off hosts can subsequently be powered back
    on to handle workload increases.
\end{itemize}
A number of cloud resource management systems, including
VMware Distributed Resource Scheduler (DRS),
Microsoft System Center, 
and Xenserver~\cite{xenserver2013} provide such functionality,
with proposals to include load balancing and power management
in OpenStack as well~\cite{wuhib2012dynamic,Beloglazov2012}.






\subsection{Powercap Distribution Examples}
\label{sec:cases}

We use several scenarios to illustrate
how \algoname can redistribute host power caps to support cloud resource management,
including enabling VM migration to correct constraint violations,
providing spare resource headroom for robustness in handling bursts, and
avoiding migrations during entitlement balancing.
In these scenarios, we assume a simple example of a cluster with two hosts.
Each host has an uncapped capacity of 2x3GHz (two CPUs, each with a 3GHz capacity)
with a corresponding peak power consumption of $600$W (values chosen for ease
of presentation).

\textbf{Enforcing constraints:} Host power caps should be
redistributed when VMs are placed initially or relocated, if necessary
to allow constraints to be respected or constraint violations to be
corrected.  For example, a cloud resource management system would move
VM(s) from a host violating affinity constraints to a target host with
sufficient capacity. However, in the case of static power cap
management, this VM movement may not be feasible because of a mismatch
between the VM reservations and the host capacity. As shown in
Figure~\ref{fig:vmmove}, host A and B have the same power cap of $480$
W, which corresponds to a power-capped capacity of $4.8$ GHz. Host A
runs two VMs, VM 1 with reservation 2.4 GHz and VM 2 with reservation
1.2 GHz.  And host B runs only one 3 GHz reservation VM.  When VM 2
needs to be colocated with VM 3 due to a new VM-VM affinity rule
between the two VMs, no target host in the cluster has sufficient
power-capped capacity to respect their combined reservations.
However, if \algoname redistributes the power caps of host A and B as
$3.6$ GHz and $6$ GHz respectively, then VM 2 can successfully be
moved by the cloud resource management system to host B to resolve the
rule violation in the cluster. Note that host A's capacity cannot be
reduced below 3.6 GHz until VM 1's migration to host B is complete or
else the reservations on host A would be violated.

\textbf{Enhancing robustness to demand bursts:} Even when VM moves do
not require changes in the host power caps, redistributing the power
caps can still benefit the robustness of the hosts to handling VM
demand bursts.  For example, as shown in Figure~\ref{fig:vmrobust},
suppose as in the previous example that VM 1 needs to move from host A
to host B because of a rule. In this case, a cloud resource management
system can move VM 1 to host B while respecting the VMs' reservations.
However, after the migration of VM 1, the \emph{headroom} between the
power capped capacity and VMs' reservations is only $0.6$ GHz on host B,
compared with $2.4$ GHz on host A. Hence, host B can only accommodate
as high as a $15\%$ workload burst without hitting the power cap while
host A can accommodate $100\%$, that is, host B is more likely to
introduce a performance bottleneck than host A. To handle this
imbalance of robustness between the two hosts, \algoname can redistribute
the power caps of host A and B as $3.6$ GHz and $6$ GHz
respectively. Now both hosts have essentially the same robustness in
term of \emph{headroom} to accommodate workload bursts.

\textbf{Reduce overhead of  VM migration:} Before entitlement balancing,
power caps should be redistributed to reduce the need for VM
migrations. Load balancing of the resources to which the VMs on a host
are entitled is a core component of cloud resource management since it
can avoid performance bottlenecks and improve system-wide
throughput. However, some recommendations to migrate VMs for load
balancing among hosts are unnecessary, given that power caps can be
redistributed to \emph{balance} workload, as shown in
Fig~\ref{fig:lb}. In this example, the VM on Host A has an entitlement
of $1.8$ GHz while the VMs on host B have a total entitlement of 3.6
GHz. The difference in entitlements between host A and B are high
enough to trigger entitlement balancing, in which VM 3 is moved from host B
to host A. After entitlement balancing, host A and B have entitlements of $3$
GHz and $2.4$ GHz respectively, that is, the workloads of both hosts
are more balanced. However, VM migration has an overhead cost and
latency related to copying the VM's CPU context and in-memory state
between the hosts involved~\cite{strunk2012}, whereas changing a host
power cap involves issuing a simple baseboard management system
command which completes in less than one
millisecond~\cite{HPDPCHowItWorks}.  \algoname can perform the cheaper
action of redistributing the power caps of hosts A and B, increasing
host B's power capped capacity to $6$ GHz after decreasing host A's
power capped capacity to $3.6$ GHz, which also results in more
balanced entitlements for host A and B. In general, the redistribution of power
caps before entitlement balancing, called \emph{powercap based entitlement balancing}, can
reduce or eliminate the overhead associated with VM migration for load
balancing, while introducing no compromise in the ability of the hosts
involved to satisfy the VMs' resource entitlements.  We note that the
goal of entitlement balancing is not \emph{absolute} balance of workload
among hosts, which may not be possible or even worthwhile given VM
demand variability, but rather reducing the imbalance of hosts'
entitlements below a predefined threshold~\cite{Gulati2011}.

\textbf{Adapting to host power on/off:} Power caps should be
redistributed when cloud resource management powers on/off host(s) to
improve cluster efficiency.  A cloud resource management system
detects when there is ongoing under-utilization of cluster host
resources leading to \emph{power-inefficiency} due to the high host
idle power consumption, and it consolidates workloads onto fewer hosts
and powers the excess hosts off.  In the example shown in
Figure~\ref{fig:poweroff}, host B can be powered off after VM 2 is
migrated to host A.  However, after host B is powered-off, it does not
consume power and hence does not need its power cap.  And the
utilization of host A is increased due to migrated VM 2, which impacts
the capacity \emph{headroom} of host A. Power cap redistribution after
powering off host B can increase the power cap of host A to $6$ GHz,
allowing the \emph{headroom} of host A to increase to $3$ GHz and
hence increase system robustness and reduce the likelihood of resource
throttling. Similarly, powercap redistribution can improve robustness
when resource management powers on hosts.

On the other hand, if there are \emph{overloaded} hosts in the
cluster, cloud resource management powers on stand-by hosts to avoid performance bottlenecks
as seen in Figure~\ref{fig:poweron}.  Due to dynamic power cap
management, active hosts can fully utilize the cluster power cap for
robustness. So a host to be powered-on may not have enough power cap to run
VMs migrated to it with suitable robustness.  \algoname can handle this
issue by redistributing the power cap among the active hosts and the
host exiting standby appropriately. For example, as shown in
Figure~\ref{fig:poweron}, host B is powered on because of the
high utilization of host A, and can only acquire $3.6$ GHz power-capped
capacity due to the limit of the cluster power budget. If VM 2 migrates
to the host B to offload the heavy usage of host A, the
\emph{headroom} of the host B will only be $1.2$ GHz, contrasting to
the \emph{headroom} of host A, which is $3.6$ GHz. However, after power cap
redistribution, the power caps of host A and B can be assigned to $4.8$ GHz
respectively, balancing the robustness of both hosts.

\section{\algoname Design}
\label{sec:operation}
In this section, we first present the design principles of \algoname.
We then give an overview of the operation of \algoname.

\subsection{\algoname Design Priciples} \label{sec:algo} 

\algoname is designed to provide power budget management to existing
resource management systems, in such a way as to support and reinforce
such systems' design and operation.  Such resource management systems
are designed to satisfy VMs' resource entitlements subject to a set of
constraints, while providing balanced headroom for demand increases
and, optionally, reduced power consumption.  \algoname improves the
operation of resource management systems, via power cap allocation
targeted to their operation.

Existing resource management systems typically involve nontrivial
complexity.  Fundamentally reimplementing them to handle hosts of
varying capacity due to power caps would be difficult and the benefit
of doing so is unclear, given the coarse-grained scales at which cloud
resource management systems operate.  In CloudPowerCap, we take the
practical approach of introducing power budget management as a
separate manager that coordinates with an existing resource management
system such that the existing system works on hosts of \emph{fixed}
capacity, with specific points at which that capacity may be modified
by \algoname in accordance with the existing system's operational
phase. Our approach therefore enhances modularity by separating power
cap and resource management, while coordinating them effectively
through well defined interfaces and protocols, as described below.


\algoname is designed to work with a cloud resource management system
with the attributes described in Section~\ref{sec:resmgmt}. Since the aim of \algoname is to enforce the cluster power budget
while dynamically managing hosts' power caps by closely coordinating
with the cloud resource management system,  \algoname
consists of three components, as shown in Figure~\ref{fig:wf},
corresponding to the three major functions of the cloud resource
management system. The three components, corresponding to main
components in DRS, execute step by step and work on two-way
interaction with components in DRS. 

\tikzstyle{materia}=[draw, fill=blue!5, text width=6.0em, text centered,
  minimum height=1.5em]
\tikzstyle{part} = [materia, text width=10em, minimum width=10em,
  minimum height=3em, rounded corners, very thick, draw=gray]
\tikzstyle{texto} = [above, text width=10em, text centered]
\tikzstyle{linepart} = [draw, very thick, color=blue!70,
 <->,  >=stealth]
\tikzstyle{line} = [draw, ultra thick, color=red!70, -latex', line
width = 2.0pt]
\tikzstyle{ur}=[draw, text centered, minimum height=0.01em]
\tikzstyle{powerline} = [draw, very thick, color=red]
\tikzstyle{mapline} = [draw, thick, color=black!50, dashed]
 
\newcommand{\blockdist}{1.3}
\newcommand{\edgedist}{1.5}
\newcommand{\step}[2]{node (p#1) [part]
  {#2}}
\newcommand{\module}[4]{%
    \path (#1.west |- #2.north)+(-0.5,0.15) node (a1) {};
    \path (#3.east |- #4.south)+(+0.5,-0.15) node (a2) {};
    \path[rounded corners, draw=black!50, very thick, fill=white,
    drop shadow]
      (a1) rectangle (a2);}

\newcommand{\background}[5]{%
    \path (#1.west |- #2.north)+(-0.25,0.35) node (a1) {};
    \path (#3.east |- #4.south)+(+0.25,-0.35) node (a2) {};
    \path[rounded corners, draw=magenta!80, dashed, very  thick]
      (a1) rectangle (a2);
    \path (a1.east |- a1.south)+(2.0,0.2) node (u1)[texto]
      {\textit{#5}};}

\begin{figure}[htbp]
    \centering
    \begin{tikzpicture}[scale=0.6,transform shape, font=\sffamily] 
      \path \step {1} {Constraints Correction};
      \path (p1.west)+(7.0, 0.0) \step{2}{Powercap Allocation};
      \path (p1.south)+(0.0,-1.6) \step{3}{Entitlement Balancing };
      \path (p3.west)+(7.0, 0.0) \step{4}{Powercap-based Entitlement Balancing};
      \path [linepart] (p4.west) --  (p3.east);
      \path (p3.south)+(0.0,-1.6) \step{5}{Dynamic Power Management};
      \path (p5.west)+(7.0, 0.0) \step{6}{Powercap Redistribution};
      \path [linepart] (p6.west) --  (p5.east);
      \path (p5.south) + (4.0, -1) node (p9) []{};
      \path (p5.south) + (6.0, -1) node (p10) []{Interaction};
      \path (p5.south) + (-2.0, -1) node (p11) []{};
      \path (p5.south) + (0, -1) node (p12) []{Workflow};
      \path [line] (p11) -- node [left] {} (p12);
      \path [linepart] (p9) -- node [left] {}(p10);

      \begin{pgfonlayer}{background}
        \module{p1}{p1}{p2}{p2}
        \module{p3}{p3}{p4}{p4}
        \module{p5}{p5}{p6}{p6}
    \end{pgfonlayer}      
      \begin{pgfonlayer}{foreground}
        \background{p1}{p1}{p5}{p5}{\textbf{DRS}}
        \background{p2}{p2}{p6}{p6}{\textbf{CloudPowerCap}}
    \end{pgfonlayer}      
        \path (p1.south) + (2.6, -0.0) node (m1) []{};
        \path (p3.north) + (2.6, 0.0) node (m2) []{};
        \path [line]  (m1.south)  --  (m2.north);
        \path (p3.south) + (2.6, -0) node (m3) []{};
        \path (p5.north) + (2.6, 0) node (m4) []{};
        \path [line]  (m3.south)  --  (m4.north);
        \path (p5.south) + (2.6, -0) node (m5) []{};
    \end{tikzpicture}
    \caption{Structure and two-way interaction of  \algoname working
      with DRS and DPM.}\label{fig:wf}
\end{figure}
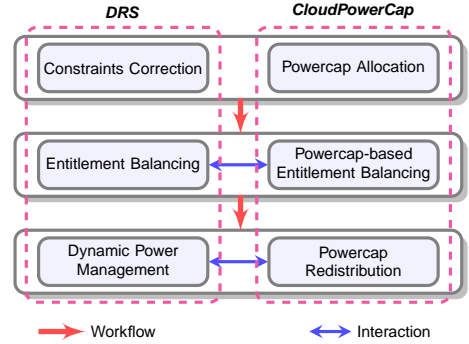

\textbf{Powercap Allocation:} During the powercap
allocation phase, potential resource management constraint correction
moves may require redistribution of host power caps.  Because
\algoname can redistribute the host power caps, the cloud resource
management system is able to correct more constraint violations than would be
possible with statically-set host power caps.

\textbf{Powercap-based Entitlement Balancing:} If the resource
management system detects entitlement imbalance over the user-set
threshold, powercap based entitlement balancing first tries to reduce
the imbalance, by redistributing power caps without actually migrating
VMs between hosts.  This is valuable because redistributing power
caps, which takes less than 1 millisecond~\cite{HPDPCHowItWorks}, is
cheaper than VM live migration in terms of overhead.  VM live
migration engenders CPU and memory overhead on both the source and
target hosts to send the VM's virtual device state, to update its
external device connections, to copy its memory one or more times to
the target host while tracing the memory to detect any writes
requiring recopy, and to make the final switchover \cite{vMotion}.
While the migration cost may be transparent to the VMs if there is
sufficient host headroom, reducing or avoiding the cost when possible
increases efficiency. Powercap Balancing may not be able to fully
address imbalance due to inherent physical host capacity limits.  If
powercap balancing cannot reduce the imbalance below the imbalance
threshold, the resource management entitlement balancing can address
the remaining imbalance by VM migration.

\textbf{Powercap Redistribution:} If the resource management
system powers on a host to match a change in workload
demands or other requirements, \algoname performs a two-pass power cap
redistribution.  First it attempts to re-allocate sufficient power cap
for that host to power-on.  If that is successful and if the system
selects the host in question after its power-on evaluation,
then \algoname redistributes the cluster power cap across the updated
hosts, to address any unfairness in the resulting power cap
distribution.  Similarly, if the system powers off a host,
its powercap can be redistributed fairly to the remaining hosts after
the host power-off operation.

\section{\algoname Implementation} \label{sec:overview}

We implemented \algoname to work with the VMware Distributed Resource
Scheduler (DRS)~\cite{drs-whitepaper} along with its optional Distributed
Power Management (DPM)~\cite{dpm-whitepaper} feature, though as we
noted in Section~\ref{sec:resmgmt}, \algoname could also complement some
other distributed resource management systems for virtualization environments.
 In this section, we first present an
overview of DRS and then detail the design of each CloudPowerCap
component and its interaction with its corresponding DRS component.


\subsection{DRS Overview}\label{sec:drs}
VMware DRS performs resource management for a cluster of ESX
hypervisor hosts. It implements the features outlined in
Section~\ref{sec:resmgmt}. By default, DRS is invoked every five
minutes.  It evaluates the state of the cluster and considers
recommendations to improve that state by executing those
recommendations in a what-if mode on an internal representation of the
cluster.  At the end of each invocation, DRS issues zero or more
recommendations for execution on the actual cluster.


At the beginning of each DRS invocation, DRS runs a phase to generate recommendations
to correct any cluster constraint violations by migrating VMs
between hosts.  Examples of such corrections include evacuating hosts
that the user has requested to enter maintenance or standby mode and
ensuring VMs respect user-defined affinity and anti-affinity business rules.
Constraint correction aims to create a
\emph{constraint compliant snapshot} of the cluster for further DRS
processing. 

DRS next performs entitlement balancing. DRS employs \emph{normalized
  entitlement} as the load metric of each host.  Denoted by $N_h$,
\emph{normalized entitlement} is defined as the sum of the per-VM
entitlements $E_i$ for each VM running on the host $h$, divided by the
capacity of the host, $C_h$, i.e., $N_h = \frac{\sum E_i}{C_h}$.
DRS's entitlement balancing algorithm uses a greedy hill-climbing
technique with the aim of minimizing the overall cluster entitlement
imbalance (i.e., the standard deviation of the hosts' normalized
entitlements).  DRS chooses as each successive move the one that
reduces imbalance most, subject to a risk-cost-benefit filter on the
move.  The risk-cost-benefit filter considers workload stability risk
and VM migration cost versus the increased balance benefit given the
last 60 minutes of VM demand history.  The move-selection step repeats
until either the load imbalance is below a user-set threshold, no
beneficial moves remain, or the number of moves generated in the
current pass hits a configurable limit based on an estimate of the
number of moves that can be executed in five minutes.

DRS then optionally runs DPM, which opportunistically saves power by
dynamically right-sizing cluster capacity to match recent workload demand,
while respecting the cluster constraints and resource controls. DPM recommends 
evacuating and powering off host(s) if the cluster contains sufficient spare resources,
and powering on host(s) if either resource demand increases appropriately or
more resources are needed to meet cluster constraints.


\newcommand{\HPC}{host power cap\xspace}
\newcommand{\HPCs}{host power caps\xspace}
\newcommand{\CPB}{cluster power budget\xspace}
\newcommand{\UPB}{unreserved power budget\xspace}

\subsection{Powercap Allocation}\label{sec:correction}
Powercap Allocation redistributes power caps if needed to allow DRS to
correct constraint violations.  DRS's ability to
correct constraint violations is impacted by \HPCs, which can limit
the available capacity on target hosts.  However, as shown in
Fig~\ref{fig:vmmove}, by increasing the host power cap, the DRS
algorithm can be more effective in correcting constraint violations.
Hence to aid DRS constraint correction, Powercap Allocation supports
redistributing the cluster's unreserved power budget, i.e., the
amount of power not needed to support running VMs' CPU and memory
reservations.  The \UPB represents the maximum amount of power cap
that can be redistributed to correct violations; insufficient \UPB
prevents the correction of constraint violations.


\tikzstyle{part1} = [materia, text width=8em, minimum width=10em,
  minimum height=3em, rounded corners, very thick, draw=gray]
\renewcommand{\step}[2]{node (p#1) [part1]
  {#2}}

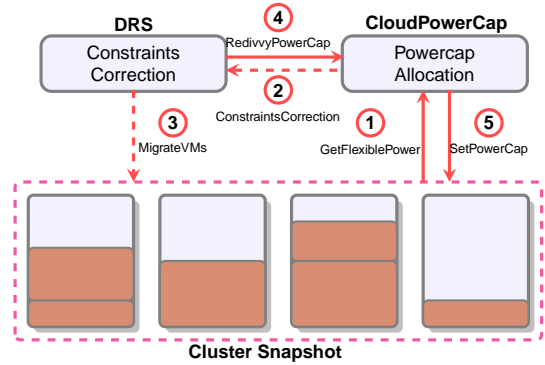
\begin{figure}[htbp]
    \centering
    \begin{tikzpicture}[scale=0.7,transform shape, font=\sffamily,
      y=0.25cm, x=0.25cm] 
      \path \step {1} {Constraints Correction};
      \path (p1.west)+(30.0, 0.0) \step{2}{Powercap Allocation};
        \draw (0, 3) node{\textbf{DRS}};
        \draw (23, 3) node{\textbf{CloudPowerCap}};
      \draw[fill=blue!5, rounded corners = 2pt, drop shadow, very
        thick, draw=gray] (-8, -20)
        rectangle (0,-10);
        \draw[fill=blue!5, rounded corners = 2pt, drop shadow, very
        thick, draw=gray] (2, -20)
        rectangle (10,-10);
      \draw[fill=blue!5, rounded corners = 2pt, drop shadow, very
        thick, draw=gray] (12, -20)
        rectangle (20,-10);
        \draw[fill=blue!5, rounded corners = 2pt, drop shadow, very
        thick, draw=gray] (22, -20)
        rectangle (30,-10);
        \draw[fill=BrickRed!50, rounded corners=2pt, thick, draw=gray] (-8, -20)
        rectangle (0,-18);
        \draw[fill=BrickRed!50, rounded corners=2pt, thick, draw=gray] (-8, -18)
        rectangle (0,-14);
        \draw[fill=BrickRed!50, rounded corners=2pt, thick, draw=gray] (2, -20)
        rectangle (10,-15);
        \draw[fill=BrickRed!50, rounded corners=2pt, thick, draw=gray] (12, -20)
        rectangle (20,-15);
        \draw[fill=BrickRed!50, rounded corners=2pt, thick, draw=gray] (12, -15)
        rectangle (20,-12);
        \draw[fill=BrickRed!50, rounded corners=2pt, thick, draw=gray] (22, -20)
        rectangle (30,-18);
        \draw[rounded corners=2pt, very thick, draw=magenta!80, dashed] (-9, -21)
        rectangle (31,-9);
        \draw (10, -22) node{\textbf{Cluster Snapshot}};
        \draw[very thick, color=red!70, <-,  >=stealth]  (22, -2.0) --
        (22, -9);
        \draw[very thick, color=red!70] (18, -4.5) circle (1);
        \draw (18, -4.5) node{\textbf{1}};
        \draw (18, -6.5) node{\scriptsize GetFlexiblePower};
        \draw[very thick, color=red!70, <-,  >=stealth, dashed]  (7,
        -0.5) -- (16, -0.5);
        \draw[very thick, color=red!70] (11, -2) circle (1);
        \draw (11, -2) node{\textbf{2}};
        \draw (11, -4) node{\scriptsize ConstraintsCorrection};
        \draw[very thick, color=red!70, ->,  >=stealth, dashed]  (0, -2.0) --
        (0, -9);
        \draw[very thick, color=red!70] (3, -4.5) circle (1);
        \draw (3, -4.5) node{\textbf{3}};
        \draw (3, -6.5) node{\scriptsize MigrateVMs};
        \draw[very thick, color=red!70, ->,  >=stealth]  (7,
        0.5) -- (16, 0.5);
        \draw[very thick, color=red!70] (11, 3.5) circle (1);
        \draw (11, 3.5) node{\textbf{4}};
        \draw (11, 1.5) node{\scriptsize RedivvyPowerCap};
        \draw[very thick, color=red!70, ->,  >=stealth]  (24, -2.0) --
        (24, -9);
        \draw[very thick, color=red!70] (27, -4.5) circle (1);
        \draw (27, -4.5) node{\textbf{5}};
        \draw (27, -6.5) node{\scriptsize SetPowerCap};
    \end{tikzpicture}
    \caption{Coordination between \algoname and DRS to correct
      constraints. Solid arrows indicate invocations of  \algoname
      functions while dashed arrows indicate invocations of DRS functions.}
    \label{fig:correction}
\end{figure}

\algoname and DRS work in coordination, as shown in
Figure~\ref{fig:correction}, to enhance the system's capability to
correct constraints violations. 
\renewcommand{\labelenumi}{\textbf{\theenumi)}}
\begin{enumerate}\itemsep1pt \parskip0pt \parsep0pt \listparindent0pt
\item Powercap Allocation first calls \emph{GetFlexiblePower} to
get \emph{flexiblePower}, which is a special clone of the current
cluster snapshot in which each host's \HPC is set to its reserved
power cap, i.e., the minimum power cap needed to support the
capacity corresponding to the reservations of the VMs currently
running on that host. 
\item The \emph{flexiblePower} is used as a parameter to call
  \emph{ConstraintsCorrection} function in DRS, which recommends VM
  migrations to enforce constraints and update hosts' reserved power
  caps for the new VM placements after the recommended
  migrations. Then DRS generates an action plan for migrating VMs.
\item As a result of performing \emph{ConstraintsCorrection}, DRS generates
  VM migration actions to correct constraints. Note that when applying
  VMs migration actions on hosts in the cluster, dependencies are
  respected between these actions and any prerequisite power cap
  setting actions generated by \algoname.
\item If some constraints are
corrected by DRS, the power caps of source and target hosts may need to be
reallocated to ensure fairness. For this case, \emph{RedivvyPowerCap} of
\algoname is called to redistribute the power cap. 
\item Finally Powercap
Allocation generates actions to set the power cap of hosts in the cluster
according to the results of \emph{RedivvyPowerCap}.
\end{enumerate}

The key function in Powercap Allocation is \emph{RedivvyPowerCap}, in
which the \UPB is redistributed after the operations for constraint
violation correction.   The inputs to this
function are S (the current snapshot of the cluster) and updated
snapshot F (after the constraint correction recommended by DRS). The
objective of RedivvyPowerCap is to distribute the cluster power budget
according to \emph{proportional resource
  sharing}~\cite{Waldspurger:1994} for maintaining fairness of
unreserved power budget distribution across hosts after the constraint
correction. The actions to change \HPC on hosts are also generated if
the hosts need more power cap than those in $S$ or less power cap
without violating VM reservation. Note these sets of power cap changes
are made appropriately dependent on the actions generated by DRS to
correct the constraint violations.


\algrenewcommand{\algorithmiccomment}[1]{\hfill$\triangleright$
  \textsf{\small#1}}
\begin{algorithm}[htbp]
\caption{Powercap Allocation\label{alg:powerred}}
\begin{algorithmic}[1]
{\small 
\Statex $S, F$: cluster snapshots before and after constraints correction;
\Statex $C_{i, S}, C_{i, F}$ power cap of the host $h_i$ in $S$ and $F$;
\Function{RedivvyPowerCap}{$S, F$}
\State $C_{needed}\gets 0$,  $C_{excess} \gets 0$
\For{each host $h_i$ in the cluster}
\If{ $ C_{i, F} > C_{i, S}$}
 \State SetPowerCap($h_i$, $C_{i, F}$)
 \State $C_{needed} \gets C_{needed} + (C_{i, F} - C_{i, S})$
\Else
 \State $C_{excess} \gets C_{excess} + (C_{i, S} - C_{i, F})$
\EndIf
\EndFor
\If{$C_{needed} > 0$}
\State $r \gets C_{needed} / C_{excess}$
\For{each host $h_i$ in the cluster}
\If{$ C_{i, F} \le C_{i, S}$}
 \State $C_{i, F} \gets C_{i, F} +  r(C_{i, S} - C_{i, F})$
 \algorithmiccomment{\small Proportional sharing}
 \State SetPowerCap($h_i$, $C_{i, F}$)
\EndIf
\EndFor
\EndIf
\EndFunction
}
\end{algorithmic}
\end{algorithm}


\def\NE{normalized entitlement }

\subsection{Entitlement Balancing}\label{sec:balance}
Entitlement balancing is critical for systems managing distributed resources,
to deliver resource entitlements and improve the responsiveness to bursts in
resource demand, and is achieved by migrating VMs between hosts. For
resource management systems like DRS without the concept of dynamic host capacity,
entitlement balancing achieves both of these goals by reducing imbalance via
migrating VMs between hosts. However, with dynamic power cap
management, \algoname can alleviate imbalance by increasing the power
caps of heavy loaded hosts while reducing the power caps of lightly
loaded hosts rather than migrating VMs between those hosts
as shown in Figure~\ref{fig:lb}.  Considering the almost negligible
overhead of power cap reconfiguration comparing to VM migration,
Powercap-based Entitlement Balancing is preferred to DRS entitlement balancing when
the cluster is imbalanced. However, because power cap adjustment
has a limited range of operation, Powercap-based Entitlement Balancing may not fully
eliminate imbalance in the cluster. But the amount of VM migration
involved in DRS entitlement balancing can be reduced significantly.

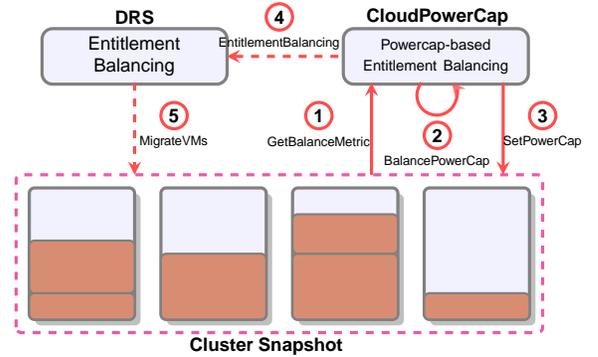
\begin{figure}[htbp]
    \centering
    \begin{tikzpicture}[scale=0.7,transform shape, font=\sffamily,
      y=0.25cm, x=0.25cm] 
      \path \step {1} {{Entitlement Balancing}};
      \path (p1.west)+(30.0, 0.0) \step{2}{\footnotesize Powercap-based Entitlement Balancing};
        \draw (0, 3) node{\textbf{DRS}};
        \draw (23, 3) node{\textbf{CloudPowerCap}};
      \draw[fill=blue!5, rounded corners = 2pt, drop shadow, very
        thick, draw=gray] (-8, -20)
        rectangle (0,-10);
        \draw[fill=blue!5, rounded corners = 2pt, drop shadow, very
        thick, draw=gray] (2, -20)
        rectangle (10,-10);
      \draw[fill=blue!5, rounded corners = 2pt, drop shadow, very
        thick, draw=gray] (12, -20)
        rectangle (20,-10);
        \draw[fill=blue!5, rounded corners = 2pt, drop shadow, very
        thick, draw=gray] (22, -20)
        rectangle (30,-10);
        \draw[fill=BrickRed!50, rounded corners=2pt, thick, draw=gray] (-8, -20)
        rectangle (0,-18);
        \draw[fill=BrickRed!50, rounded corners=2pt, thick, draw=gray] (-8, -18)
        rectangle (0,-14);
        \draw[fill=BrickRed!50, rounded corners=2pt, thick, draw=gray] (2, -20)
        rectangle (10,-15);
        \draw[fill=BrickRed!50, rounded corners=2pt, thick, draw=gray] (12, -20)
        rectangle (20,-15);
        \draw[fill=BrickRed!50, rounded corners=2pt, thick, draw=gray] (12, -15)
        rectangle (20,-12);
        \draw[fill=BrickRed!50, rounded corners=2pt, thick, draw=gray] (22, -20)
        rectangle (30,-18);
        \draw[rounded corners=2pt, very thick, draw=magenta!80, dashed] (-9, -21)
        rectangle (31,-9);
        \draw (10, -22) node{\textbf{Cluster Snapshot}};
        \draw[very thick, color=red!70, <-,  >=stealth]  (18, -2.0) --
        (18, -9);
        \draw[very thick, color=red!70] (14, -4.5) circle (1);
        \draw (14, -4.5) node{\textbf{1}};
        \draw (14, -6.5) node{\scriptsize GetBalanceMetric};
        \draw[very thick, color=red!70, ->,  >=stealth, fill opacity=0.8]  (21.8, -2.0)
        arc[x radius=1.5, y radius =1.5, start angle=140, end
        angle=400];
        \draw[very thick, color=red!70] (23, -6.0) circle (1);
        \draw (23, -6.0) node{\textbf{2}};
        \draw (23, -8) node{\scriptsize BalancePowerCap};
        \draw[very thick, color=red!70, ->,  >=stealth]  (28, -2.0) --
        (28, -9);
        \draw[very thick, color=red!70] (31, -4.5) circle (1);
        \draw (31, -4.5) node{\textbf{3}};
        \draw (31, -6.5) node{\scriptsize SetPowerCap};
        \draw[very thick, color=red!70, <-,  >=stealth, dashed]  (7,
        -0) -- (16, -0);
        \draw[very thick, color=red!70] (11, 3) circle (1);
        \draw (11, 3) node{\textbf{4}};
        \draw (11, 1) node{\scriptsize EntitlementBalancing};
        \draw[very thick, color=red!70, ->,  >=stealth, dashed]  (0, -2.0) --
        (0, -9);
        \draw[very thick, color=red!70] (3, -4.5) circle (1);
        \draw (3, -4.5) node{\textbf{5}};
        \draw (3, -6.5) node{\scriptsize MigrateVMs};
    \end{tikzpicture}
    \caption{Work flow of Powercap-based Entitlement Balancing and its interaction with
      DRS entitlement balancing.  Solid arrows indicate invocations of \algoname
      functions while dashed arrows indicate invocations of DRS functions.}
    \label{fig:powercapbalancing}
\end{figure}

The process of powercap based entitlement balancing and its interaction with DRS load
balancing are shown in Figure~\ref{fig:powercapbalancing}.

\algrenewcommand{\algorithmiccomment}[1]{\hfill$\triangleright$ \textsf{\small#1}}
 \begin{algorithm}[htbp]
 \caption{Powercap-based Entitlement Balancing} \label{alg:balance}
 \begin{algorithmic}[1]
{\small 
\Statex $S, F$:  cluster snapshot before and after Powercap  Based
Entitlement Balancing
\Statex $h, l$: hosts with highest and lowest normalized
entitlement
\Statex $\hat{C}_i$: peak capacity of the host $i$
\Statex $\bar{C}_i$ :
capacity of the host $i$ corresponding to average normalized entitlement of the cluster
\Function{BalancePowerCap}{S}
\State $F \gets S$,  pcBal $\gets$ false
\While{ Cluster is imbalanced}
\State Choose $h$ and $l$ from the cluster
 \State $C_{needed} \gets min(\hat{C}_h,
\bar{C}_h) - C_h$
 \State $C_{avail} \gets  C_l- \bar{C}_l$
 \If{ $C_{needed} = 0$ or $C_{avail} = 0$} 
 \State break \algorithmiccomment{Then invoke DRS 
   entitlement balancing}
 \Else
 \State pcBal $\gets$ true
 \EndIf
\State Add $C_{avail}$ to $h$ and reduce $C_{needed}$ from $l$
 \State Recompute cluster balance metric on $F$
\EndWhile
\If{pcBal = true}
 \State Set power cap of hosts according to $F$
\EndIf
\State return $F$ 
\EndFunction
}
 \end{algorithmic}
 \end{algorithm}

\newcommand*\circled[1]{\tikz[baseline=(char.base)]{
            \node[shape=circle,draw,inner sep=1pt] (char) {#1};}}
\renewcommand{\labelenumi}{\textbf{\theenumi})}
\begin{enumerate}\itemsep1pt \parskip0pt \parsep0pt \listparindent0pt
\item To acquire the status of entitlement imbalance of the cluster,
  Powercap-based Entitlement Balancing first calculates the DRS imbalance metric for
  the cluster (i.e., the standard deviation of the hosts' normalized
  entitlements).
\item Then Powercap-based Entitlement Balancing tries to reduce the entitlement
  imbalance among hosts by adjusting their power caps in
  accordance with their normalized entitlements.
\item If Powercap-based Entitlement Balancing is able to impact cluster imbalance, its
  \HPC redistribution actions are added to the recommendation list, with the \HPC
  reduction actions being prerequisites of the increase actions.
\item If Powercap-based Entitlement Balancing has not fully balanced the entitlement
  among the hosts, DRS entitlement balancing is invoked on the results
  of Powercap-based Entitlement Balancing to reduce entitlement imbalance further.
\item DRS may generate actions to migrate VMs.
\end{enumerate}



The sketch of the key function \emph{BalancePowerCap} in Powercap-base Entitlement
Balancing is shown in Algorithm~\ref{alg:balance}, which was developed
along the lines of \emph{progressive filling} to achieve max-min
fairness~\cite{Bertsekas1992}.  The algorithm progressively increases
the \HPC of the host(s) with highest \NE while progressively reducing
the \HPC of the host(s) with lowest \NE.  This process is repeated
until either the DRS imbalance metric crosses the balance threshold or
any of the host(s) with highest \NE reach their peak capacity and
hence further reduction in overall imbalance is limited by those
hosts.

\subsection{Powercap Redistribution}
Powercap Redistribution responds to DPM dynamically
powering on/off hosts. When CPU or memory utilization becomes high, DPM
recommends powering on hosts and redistributing the VMs across the
hosts to reduce per-host load.  Before the host is powered on,
Powercap Redistribution ensures that sufficient power cap is assigned to
the powering-on host. On the other hand, when both CPU and memory
utilization are low for a sustained period, DPM may recommend
consolidating VMs onto fewer hosts and powering off the remaining hosts
to save energy.  In this case, Powercap Redistribution distributes the
power caps of the powered-off hosts among the active hosts to increase
their capacity.



\begin{figure}[H]
    \centering
    \begin{tikzpicture}[scale=0.7,transform shape, font=\sffamily,
      y=0.25cm, x=0.25cm] 
      \path \step {1} {DPM};
      \path (p1.west)+(30.0, 0.0) \step{2}{Powercap Redistribution};
        \draw (0, 3) node{\textbf{DRS}};
        \draw (23, 3) node{\textbf{CloudPowerCap}};
      \draw[fill=blue!5, rounded corners = 2pt, drop shadow, very
        thick, draw=gray] (-8, -20)
        rectangle (0,-10);
        \draw[fill=blue!5, rounded corners = 2pt, drop shadow, very
        thick, draw=gray] (2, -20)
        rectangle (10,-10);
      \draw[fill=blue!5, rounded corners = 2pt, drop shadow, very
        thick, draw=gray] (12, -20)
        rectangle (20,-10);
        \draw[fill=blue!5, rounded corners = 2pt, drop shadow, very
        thick, draw=gray] (22, -20)
        rectangle (30,-10);
        \draw[fill=BrickRed!50, rounded corners=2pt, thick, draw=gray] (-8, -20)
        rectangle (0,-18);
        \draw[fill=BrickRed!50, rounded corners=2pt, thick, draw=gray] (-8, -18)
        rectangle (0,-14);
        \draw[fill=BrickRed!50, rounded corners=2pt, thick, draw=gray] (2, -20)
        rectangle (10,-15);
        \draw[fill=BrickRed!50, rounded corners=2pt, thick, draw=gray] (12, -20)
        rectangle (20,-15);
        \draw[fill=BrickRed!50, rounded corners=2pt, thick, draw=gray] (12, -15)
        rectangle (20,-12);
        \draw[fill=BrickRed!50, rounded corners=2pt, thick, draw=gray] (22, -20)
        rectangle (30,-18);
        \draw[rounded corners=2pt, very thick, draw=magenta!80, dashed] (-9, -21)
        rectangle (31,-9);
        \draw (10, -22) node{\textbf{Cluster Snapshot}};
        \draw[very thick, color=red!70, <-,  >=stealth, dashed]  (-2, -2.0) --
        (-2, -9);
        \draw[very thick, color=red!70] (-5, -4.5) circle (1);
        \draw (-5, -4.5) node{\textbf{1}};
        \draw (-5, -6.5) node{\scriptsize GetUtilization};
        \draw[very thick, color=red!70, ->,  >=stealth]  (7,
        0.5) -- (16, 0.5);
        \draw[very thick, color=red!70] (11, 3.5) circle (1);
        \draw (11, 3.5) node{\textbf{2}};
        \draw (11, 1.5) node{\scriptsize RedistributePowerCap};
        \draw[very thick, color=red!70, <-,  >=stealth, dashed]  (7,
        -0.5) -- (16, -0.5);
        \draw[very thick, color=red!70] (11, -2) circle (1);
        \draw (11, -2) node{\textbf{3}};
        \draw (11, -4) node{\scriptsize TryPowerOnHost};
        \draw[very thick, color=red!70, ->,  >=stealth]  (22, -2.0) --
        (22, -9);
        \draw[very thick, color=red!70] (25, -4.5) circle (1);
        \draw (25, -4.5) node{\textbf{5}};
        \draw (25, -6.5) node{\scriptsize SetPowerCap};
        \draw[very thick, color=red!70, ->,  >=stealth, dashed]  (2, -2.0) --
        (2, -9);
        \draw[very thick, color=red!70] (5, -4.5) circle (1);
        \draw (5, -4.5) node{\textbf{4}};
        \draw (5, -6.5) node{\scriptsize PowerOnHosts};
    \end{tikzpicture}
    \caption{Coordination between \algoname and DRS and DPM in
      response to power on/off hosts. Solid arrows indicate to invoke \algoname
      functions while dashed arrows indicate to invoke DRS functions.}
    \label{fig:powercapdpm}
\end{figure}
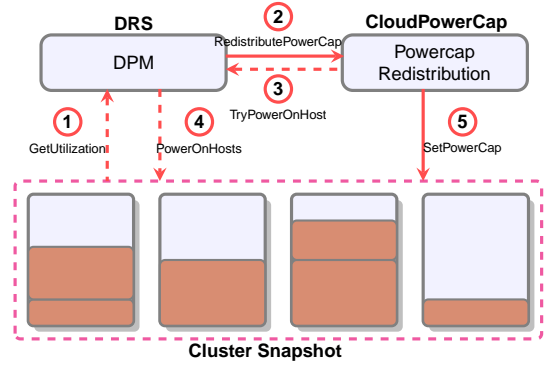

\def\UCP{unreserved cluster power budget } The coordination between
Powercap Redistribution and DPM when DPM attempts to power on a host
is depicted in Figure~\ref{fig:powercapdpm}.
\renewcommand{\labelenumi}{\textbf{\theenumi})}
\begin{enumerate}\itemsep1pt \parskip0pt \parsep0pt \listparindent0pt
\item If there is sufficient \UCP to set the target host's power cap
  to peak, the host obtains its peak \HPC from the \UCP and no power
  cap redistribution is needed.
\item If the current \UCP is not sufficient,
  \emph{RedistributePowerCap} is invoked to allow the powering-on candidate host
  to acquire more power from those hosts with lower CPU utilization.
\item DPM decides whether to power on the candidate host given its
  updated power cap after redistribution and its ability to reduce host
  high utilization in the cluster.
\item If the host is chosen for power-on, the normal DPM function is
  invoked to generate the action plan for powering on the host.
\item If DPM decides to recommend the candidate power-on, any needed \HPC
  changes are recommended as prerequisites to the host power-on.
\end{enumerate}

The algorithm of redistributing power caps is straightforward.  To
acquire sufficient power caps to power on a host, the hosts with lower
utilization have their power caps reduced under the constraint of not
causing those hosts to enter the high utilization range that would
trigger DPM to power on another host.

When a host is being considered for powering-off, the portion of its \HPC
currently above its utilization could be made available for
redistribution to other powered-on hosts whose \HPCs are below peak,
providing more target capacity for evacuating VMs.

\subsection{Implementation Details}

We implemented \algoname on top of VMware's production version of DRS.
Like DRS, \algoname is written in C++.  The entire implementation of
\algoname comprises less than 500 lines of C++ code, which
demonstrates the advantage of instantiating power budget management as
a separate module that coordinates with an existing resource manager
through well-defined interfaces.

As described previously in this section, DRS operates on a snapshot
of the VM and host inventory it is managing.  The main change we made for
DRS to interface with \algoname was to enhance the DRS method for determining a host's
CPU capacity to reflect the host's current power cap setting in the snapshot.
Other small changes were made to support the \algoname functionality,
including specifying the power budget, introducing a new action that DRS could
issue for changing a host's power cap, and providing support for testability.

During \algoname initialization, for each host, the mapping between
its current power cap and its effective capacity is established by the
mechanisms described in Section~\ref{sec:power_model}.  For a
powered-on host, the power cap value should be in the range between
the host's idle and peak power.  
When computing power-capped capacity of a host based on the power
model~\eqref{eq:power_model}, it is important to ensure that
the capacity reserved by the hypervisor on the host is fully
respected.  Hence, the power-capped capacity $C_{mcapped}$ managed by the resource
management system, i.e., managed capacity, is computed as:
\begin{equation}
  C_{mcapped} = C_{capped} - C_{H},
\end{equation}
where the power-capped raw capacity $C_{capped}$ is computed using
Equation~\eqref{eq:linpower} and $C_H$ is the capacity reserved by the
hypervisor.

The implementation of
Powercap Allocation entailed updating corresponding DRS methods to understand
that a host's effective capacity available for constraint correction could be
increased using the unreserved power budget, and adding a powercap redivvy step
optionally run at the end of the constraint correction step.
Powercap Balancing, which leverages elements of the powercap redivvying code,
involved creating a new method to be called before the DRS balancing method.
Powercap Redistribution changed DPM functions to consider whether to turn on/off
hosts based not only on utilization but also on the available power budget.


\section{Evaluation}
\label{sec:implementation}

In this section, we evaluate CloudPowerCap in the DRS simulator under
three interesting scenarios.  The first experiment evaluates
\algoname's capability to rebalance normalized entitlement among hosts
while avoiding the overhead of VM migration.  The second experiment
shows \algoname reallocates the power budget of a powered-off host to
allow hosts to handle demand bursts.  The third experiment shows how
\algoname allows CPU and memory capacity trade-offs to be made at
runtime.  This experiment includes a relatively large host inventory
to show the capacity trade-offs at scale.

In these experiments, we compare \algoname against two baseline
approaches of power cap management: \emph{StaticHigh} and
\emph{Static}. Both approaches assign equal power cap to each host in
the cluster at the beginning and maintain those power caps throughout
the experiment.  \emph{StaticHigh} sets power cap of the host
to its peak power, maximizing throughput of CPU intensive
applications. However for applications in which memory or storage
become constrained resources, it can be beneficial to support more
servers to provision more memory and storage.  Hence in \emph{Static},
the power cap of a host is intentionally set lower than the peak
power of the host. Compared with \emph{StaticHigh}, more servers may be
placed with \emph{Static} to enhance the throughput of applications with
memory or storage as constrained resources. However both approaches
lack the capability of flexible power cap allocation to
respond to workload spikes and demand variation.

\subsection{DRS Simulator}

The DRS simulator~\cite{drsdesign} is used in developing and testing all DRS algorithm
features.  It provides a realistic execution environment, while
allowing much more flexibility and precision in specifying VM demand
workloads and obtaining repeatable results than running on real
hardware. 

The DRS simulator simulates a cluster of ESX hosts and VMs. A host can be
defined using parameters including number of physical cores, CPU
capacity per core, total memory size, and power consumption at idle
and peak.  A VM can be defined in terms of number of configured
virtual CPUs (vCPUs) and memory size.  Each VM's workload can be
described by an arbitrary function over time, with the simulator
generating CPU and memory demand for that VM based on the
specification.

Given the input characteristics of ESX hosts and the VMs' resource
demands and specifications, the simulator mimics ESX CPU and memory
schedulers, allocating resources to the VMs in a manner consistent
with the behavior of ESX hosts in a real DRS cluster.  The simulator
supports all the resource controls supported by the real ESX
hosts. The simulator can support vMotion of VMs, and models the cost of
vMotion and its impact on the workload running in the VM. The simulator models the ESX
hypervisor CPU and memory overheads.

The simulator is able to estimate the power consumption of the ESX
hosts based on the power model given in Equation~\eqref{eq:linpower}
in Section~\ref{sec:rackpower}.  For this work, the simulator was
updated to respect the CPU capacity impact associated with a host's
power cap.




\subsection{Headroom Rebalancing}

\begin{figure}[htbp]
\centering
   \subfloat[CloudPowerCap]{
    \label{fig:cpc}
    \includegraphics[scale=0.20]{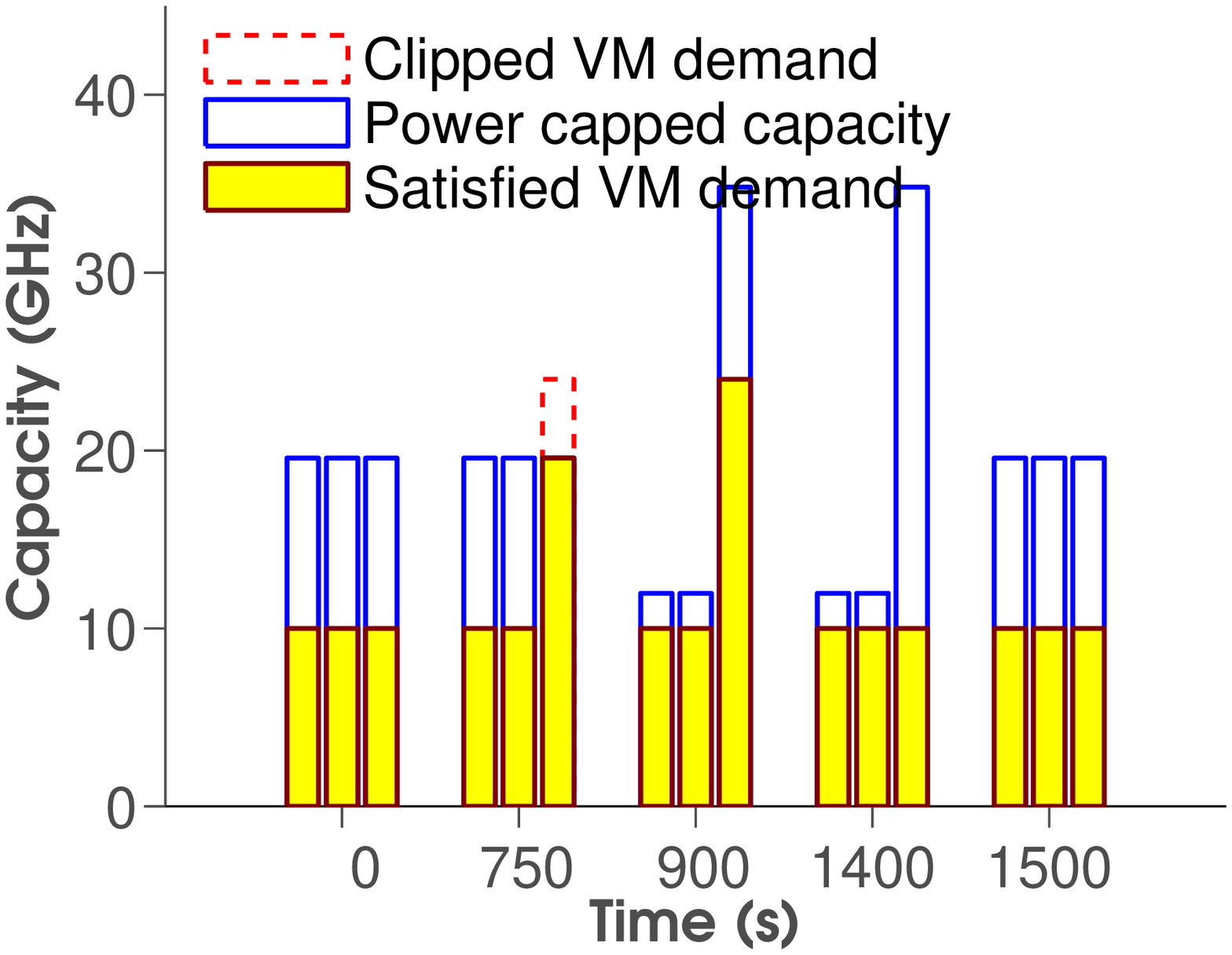}
     } 
  \subfloat[Static]{
    \label{fig:static}
      \includegraphics[scale=0.20]{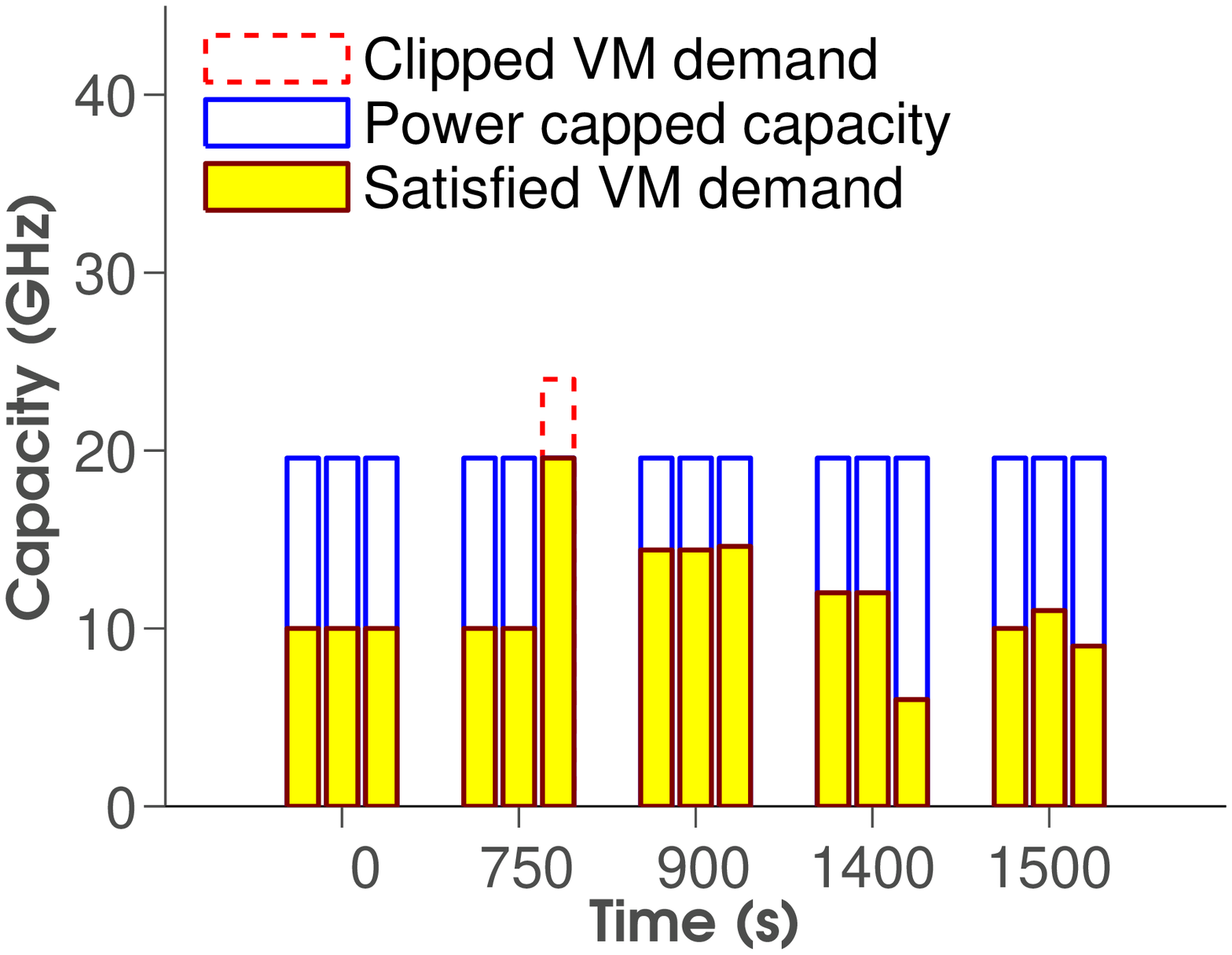} 
   }
  \vspace{-2mm}
  \caption{Headroom balancing on a group of 3 hosts. Hosts are grouped at each event time. }
  \label{fig:balance}
\end{figure}

\algoname can reassign power caps to balance headroom for bursts,
providing a quick response to workload imbalance due to VM demand
changes.  Such reassignment of power caps can improve robustness of
the cluster and reduce or avoid the overhead of VM migration for load
balancing.  To evaluate impact of \algoname on headroom balancing, we
perform an experiment in which 30 VMs, each with 1vCPU and 8GB memory,
run on 3 hosts with the configuration shown in Table~\ref{tab:server}.
Figures~\ref{fig:cpc} and~\ref{fig:static} plot the simulation results
under CloudPowerCap and Static with a static power cap allocation of
250W per host, respectively.  Initially, at time 0 seconds, the VMs
are each executing similar workloads of 1 GHz CPU and 2 GB memory
demand, and are evenly distributed across the hosts.  At time 750
seconds, the VMs on one host spike to 2.4 GHz demand, thereby
increasing the demand on that host above its power-capped capacity.
When DRS is next invoked at time 900 seconds (running every 300
seconds by default), its goal is to rebalance the hosts' normalized
entitlements.  Under the static power cap, DRS migrates the VMs to
balance the normalized entitlements.  In contrast, \algoname reassigns
the hosts' power caps to reduce the caps on the light-loaded hosts (to
215W) and increase them on the heavy-loaded host (to 320W).  This
addresses the host overutilization and imbalance without requiring
vMotion latency and overhead, which is particularly important in this
case, since the overhead further impacts the workloads running on the
overutilized host.  At time 1400 seconds, the 2.4 GHz VM demand spike ceases,
and those VMs resume running at their original 1 GHz demand until the
experiment ends at time 2100 seconds.  Again, \algoname avoids the need for
migrations by reassigning the host power caps to their original
values. In contrast, Static performs two DRS entitlement balancing
phases and migrates several VMs at time 900 seconds and 1500 seconds.

\begin{table}[ht]
\centering
{\small
\begin{tabular}{@{}ccc@{}}
\toprule
 &  \textbf{CPU Payload Ratio}   & \textbf{vMotion}   \\
\midrule
CPC       &  0.99 & 0 \\
Static    &  0.89 & 7 \\
StaticHigh &  1.00 & 0 \\
\bottomrule
\end{tabular}
}
\caption{CloudPowerCap (CPC) rebalancing without migration overhead}
\label{table:CPCbalance}
\end{table}

Table~\ref{table:CPCbalance} compares the CPU payload
delivered to the VMs under \algoname, Static using 250W static
host power caps, as well as StaticHigh using the power caps equivalent to
the peak capacity of the host.  For Static, the vMotion CPU overhead
has a significant overall impact on the CPU payload delivered to the
VMs because the host is overutilized during the burst and the cycles
needed for vMotion directly impact those available for VM use.  For
\algoname, there is a relatively small impact to performance after the burst
and before DRS can run \algoname to reallocate the host power caps.
The power cap setting operation itself can be executed by the host
within 1 millisecond and introduces minor payload overhead.

\subsection{Standby Host Power Reallocation}

\algoname can reallocate standby hosts' power cap to increase the capacity
of powered-on hosts and thereby their efficiency and ability to handle
bursts.  To demonstrate this, we consider the same initial setup in
terms of hosts and VMs as in the previous experiment.  In this case,
all VMs are running a similar workload of 1.2 GHz and 2 GB memory
demand.  At time 750 seconds, each VM's demand reduces to 400 MHz, and when
DRS is next invoked at time 900 seconds, DPM recommends that the VMs be
consolidated onto two hosts and that another host is
powered-off.  After the host has been evacuated and powered-off at
time 1200 seconds, \algoname reassigns its power cap to 0 and reallocates the
rack power budget to the two remaining hosts, setting their power caps
to 320W each.  At time 1400 seconds, there is an unexpected spike.  In the
case of statically-assigned power caps, the host that was powered-off
is powered back on to handle the spike, but in the \algoname case, the
additional CPU capacity available on the 2 remaining hosts given their
320 W power caps is sufficient to handle this spike and the
powered-off host is not needed.

\begin{table}[ht]
\centering
{\small
\begin{tabular}{@{}cccc@{}}
\toprule
  & \textbf{CPU Payload Ratio}   & \textbf{vMotion}  & \textbf{Power Ratio} \\
\midrule
CPC        & 1.00 & 10 & 1.00 \\
Static     & 0.98 & 19 & 1.36 \\
StaticHigh & 1.00 & 10 & 1.00 \\
\bottomrule
\end{tabular}
}
\caption{CloudPowerCap (CPC) reallocating standby host power}
\label{table:CPCDPM}
\end{table}

Table~\ref{table:CPCDPM} compares the CPU payload in cycles delivered
to the VMs for \algoname, Static, and StaticHigh.  In this case, a number
of additional vMotions are needed for Static, but the overhead of
these vMotions does not significantly impact the CPU payload,
because there is plenty of headroom to accomodate this overhead.  However,
Static consumes much more power than the other 2 cases, since
powering the additional host back on and repopulating it consumes
significant power.  In contrast, \algoname is able to match the power efficiency of
the baseline, by being able to use peak capacity of the powered-on
hosts.

\subsection{Flexible Resource Capacity}

\begin{figure}[htbp]
\centering
   \subfloat[CloudPowerCap]{
    \label{fig:flex_cpc}
    \includegraphics[scale=0.20]{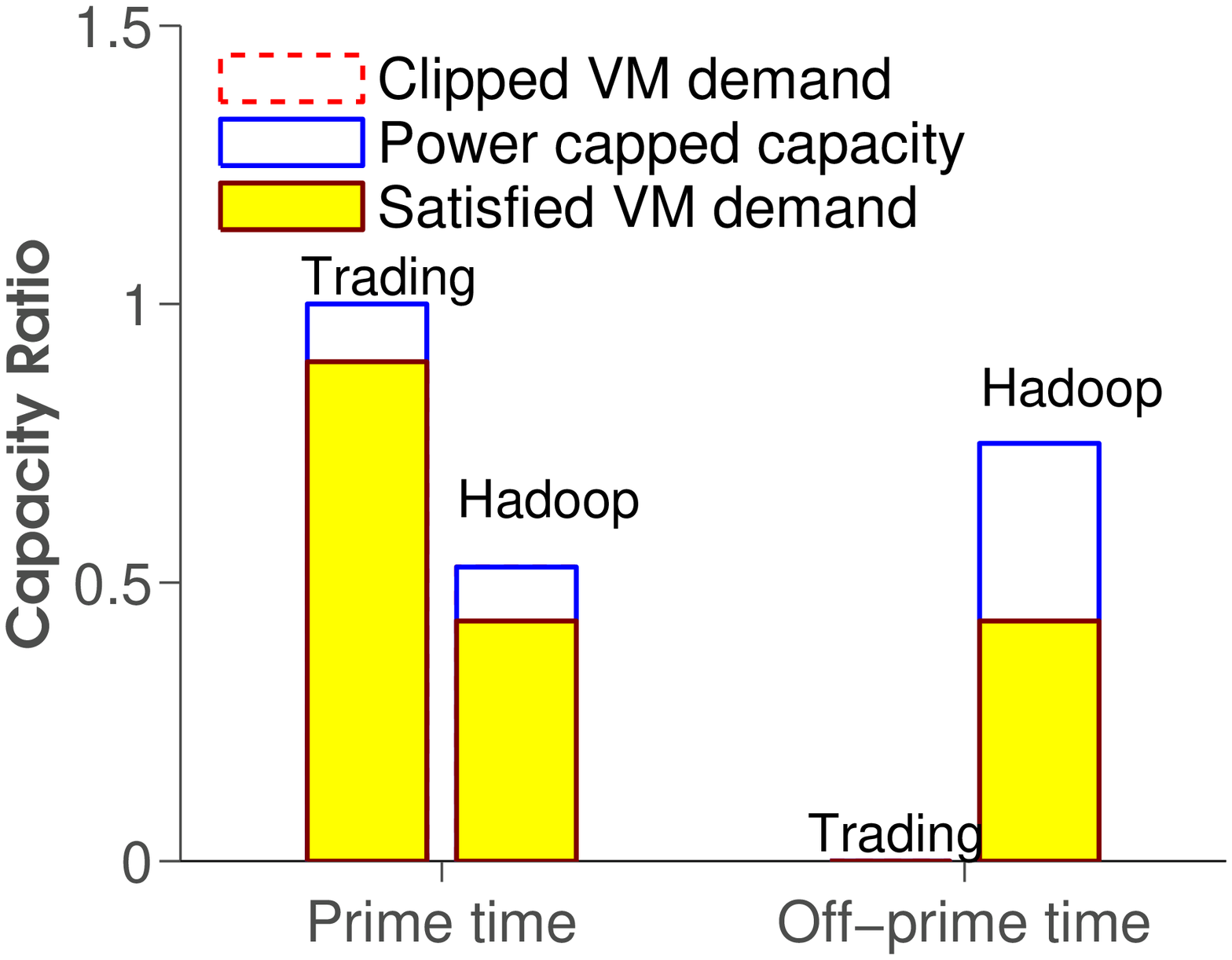}
     }
  \subfloat[Static]{
    \label{fig:flex_static}
      \includegraphics[scale=0.20]{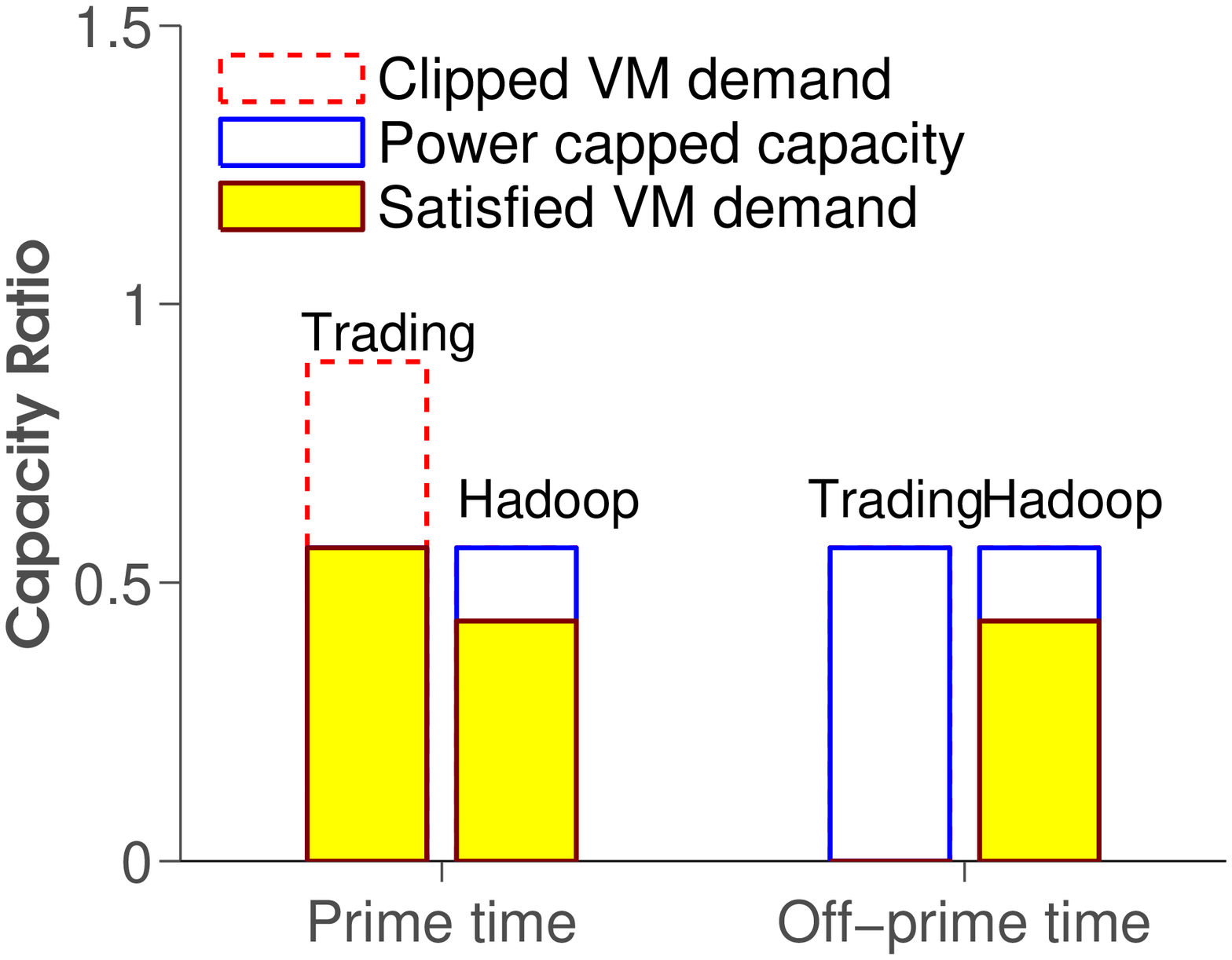} 
   }
  \vspace{-2mm}
  \caption{Trade-offs between dynamic resource capacities. \emph{Trading} indicates a group of servers running
    production trading VMs while \emph{Hadoop}
    represents servers run production Hapdoop compute VMs. }
  \label{fig:flexible}
\end{figure}

CloudPowerCap supports flexible use of power to allow trade-offs
between resource capacities to be made dynamically.  To illustrate
such a trade-off at scale, we consider a cluster of hosts as described
in Section 2.1.  We model the situation in which the cluster is used to run both production trading
VMs and production hadoop compute VMs.  The trading VMs are configured
with 2 vCPUs and 8 GB and they are idle half the day (off-prime time),
and they run heavy workloads of 2x2.6 GHz and 7 GB demand the other
half of the day (prime time).  They access high-performance shared
storage and hence are constrained to run on hosts with access to that
storage, which is only mounted on 8 hosts in the cluster.  The hadoop
compute VMs are configured with 2 vCPUs and 16 GB and each runs a
steady workload of 2x1.25 GHz and 14 GB demand.  They access local
storage and hence are constrained to run on their current hosts and
cannot be migrated.  During prime time, the 8 servers running the
trading VMs do not receive tasks for the hadoop VMs running on those
servers; this is accomplished via an elastic scheduling response to
the reduced available resources
~\cite{White2009}. Figure~\ref{fig:flexible} shows the simulation
results of the cluster under \algoname and the Static configuration of
power caps.

Table~\ref{table:CPCScale} compares the CPU and memory payload
delivered for three scenarios, and shows the impact on the trading VMs.  The staticHigh scenario involves
deploying 25 servers with power caps of 320 W, which immediately and
fully supports the trading VMs prime time demand but limits the
overall available memory and local disks in the cluster associated
with the 25 servers.  The Static scenario instead involves deploying 32
servers with each host power cap statically set to 250 Watts.  This
scenario allows more memory and local disks to be accessed, increasing
the overall CPU and memory payload delivered because more hadoop work
can be accomplished, but limits the peak CPU capacity of each host,
meaning that the trading VMs run at only 62 percent of their prime
time demand.  With CloudPowerCap, the benefits to the hadoop workload
of the static scenario are retained, but the power caps of the hosts
running the trading VMs can be dynamically increased, allowing those
VMs' full prime time demand to be satisfied.

\begin{table}[ht]
\centering
{\small
\begin{tabular}{@{}cccc@{}}
\toprule
 & \textbf{CPU Ratio}   & \textbf{Mem Ratio}   &
 \textbf{Trading Ratio} \\
\midrule
CPC        & 1.24 & 1.28 & 1.00 \\
Static     & 1.21 & 1.28 & 0.62 \\
StaticHigh & 1.00 & 1.00 & 1.00 \\
\bottomrule
\end{tabular}
}
\caption{CloudPowerCap (CPC) enabling flexible resource
  capacity. Trading ratio indicates the ratio that production
  trading VMs demands in prime time are satisfied. }
\label{table:CPCScale}
\end{table}

\section{Related Work}
\label{sec:related} 
Several research projects have considered power cap management for
virtualized infrastructure~\cite{Raghavendra2008, Nathuji2009,
  Lim2011, Wang2011,Nathuji2009a}.  Among them, the research mostly related to our work
is~\cite{Nathuji2009a}, in which authors proposed VPM tokens, an
abstraction of changeable weights, to support power budgeting in
virtualized environment. Like our work, VPM tokens enables shifting
\emph{power budget slack} which corresponds to \emph{headroom} in this
paper, between hosts. However the power cap management system based on
VPM tokens are independent of resource
management systems and may generate conflicting actions without
coordination mechanisms.

In contrast, interoperating with a cloud resource management system like DRS also
allows \algoname to support interesting additional features: 1)
CloudPowerCap accommodates consolidation of physical servers caused by
dynamic power management while previous work assumed a fixed working
server set, 2) CloudPowerCap is able to handle and facilitate VM
migration caused by correcting constraints imposed on physical servers
and VMs, 3) CloudPowerCap can also deal with and enhance power cap
management in the presence of load balancing which is not considered
in the previous paper.

The authors of~\cite{Raghavendra2008} describe managing performance
and power management goals at server, enclosure, and data center level
and propose handling the power cap hierarchically across multiple
levels.  Optimization and feedback control algorithms are employed to
coordinate the power management and performance indices for entire
clusters.  In~\cite{Wang2011}, the authors build a framework to
coordinate power and performance via Model Predictive Control through
DVFS (Dynamic Voltage and Frequency Scaling). To provide power cap
management through the VMs management layer, ~\cite{Nathuji2009}
proposed throttling VM CPU usage to respect the power cap. In their
approach, feedback control is also used to enforce the power cap while
maintaining system performance.  Similarly, the authors
in~\cite{Lim2011} also discussed data center level power cap
management by throttling VM resource
allocation. Like~\cite{Raghavendra2008}, they also adopted a
hierarchical approach to coordinate power cap and performance goals.

While all of these techniques attempt to manage both power and
performance goals, their resource models for the performance goals are
incomplete in various ways. For examples, none of the techniques
support guaranteed SLAs (reservations) and fair share scheduling
(shares).  Some build a feedback model needing application-level
performance metrics acquired from cooperative clients, which is rare
especially in public clouds~\cite{Ben-Yehuda2012}.  



\section{Conclusion}
\label{sec:summary}

Many modern data centers have underutilized racks.  Server vendors
have recently introduced support for per-host power caps, which provide a
server-enforced limit on the amount of power that the server can draw,
improving rack utilization.  However, this approach is tedious
and inflexible because it needs involvement of human operators and
does not adapt in accordance with workload variation.  This paper
presents \algoname to manage a cluster power budget for a virtualized
infrastructure. In coordination with resource management,
CloudPowerCap provides holistic and adaptive power budget management
framework to support service level agreements, fairness in spare power
allocation, entitlement balancing and constraint enforcement.

\renewcommand{\baselinestretch}{0.96}
\def\IEEEbibitemsep{0pt plus .5pt}
\bibliographystyle{IEEEtran}
\small
\bibliography{paper}

\end{document}